\documentclass[aps,prx,twocolumn,superscriptaddress]{revtex4-1}
\usepackage[colorlinks,bookmarks=false,citecolor=blue,linkcolor=red,urlcolor=blue]{hyperref}
\usepackage[utf8x]{inputenc}
\usepackage{MnSymbol}           
\usepackage{graphicx}
\usepackage{xcolor}             
\usepackage{epstopdf}
\usepackage{lipsum}

\newcommand{\aRuCl}     {$\alpha$-RuCl$_3$}

\begin{document}

\title{Spin-orbit excitation energies, anisotropic exchange, and magnetic phases of honeycomb \aRuCl}

\author{Ravi Yadav}
\affiliation{Institute for Theoretical Solid State Physics, IFW Dresden, Helmholtzstrasse 20, 01069 Dresden, Germany}
\author{Nikolay A.~Bogdanov$^{\dagger}$}
\affiliation{Institute for Theoretical Solid State Physics, IFW Dresden, Helmholtzstrasse 20, 01069 Dresden, Germany}
\AtEndDocument{\bigskip{\footnotesize%
\addvspace{\medskipamount}
{$^{\dagger}$\,\textit{Current address: Max Planck Institute for Solid State Research, Heisenbergstrasse 1, 70569 Stuttgart, Germany}} \par
}}
\author{Vamshi  M.~Katukuri$^{\ddag}$}
\affiliation{Institute for Theoretical Solid State Physics, IFW Dresden, Helmholtzstrasse 20, 01069 Dresden, Germany}
\AtEndDocument{{\footnotesize%
\addvspace{\medskipamount}
{$^{\ddag}$\,\textit{Current address: Institute of Theoretical Physics, Ecole Polytechnique F\'{e}derale de Lausanne (EPFL), CH-1015 Lausanne, Switzerland}} \par
}}
\author{Satoshi Nishimoto}
\affiliation{Institute for Theoretical Solid State Physics, IFW Dresden, Helmholtzstrasse 20, 01069 Dresden, Germany}
\affiliation{Department of Physics, Technical University Dresden, Helmholtzstrasse~10, 01069 Dresden, Germany}
\author{Jeroen van den Brink}
\affiliation{Institute for Theoretical Solid State Physics, IFW Dresden, Helmholtzstrasse 20, 01069 Dresden, Germany}
\affiliation{Department of Physics, Technical University Dresden, Helmholtzstrasse~10, 01069 Dresden, Germany}
\affiliation{Department of Physics, Harvard University, Cambridge, Massachusetts 02138, USA}
\author{Liviu Hozoi}
\affiliation{Institute for Theoretical Solid State Physics, IFW Dresden, Helmholtzstrasse 20, 01069 Dresden, Germany}

\date{\today}

\begin{abstract}
Using quantum chemistry calculations we shed fresh light on the electronic structure and magnetic
properties of \aRuCl , a proposed realization of the honeycomb Kitaev spin model.
It is found that the nearest-neighbor Kitaev exchange $K$ is weaker than in $5d^5$ Ir oxides
but still larger than other effective spin couplings.
The electronic-structure computations also indicate a ferromagnetic $K$ in the halide, which is supported by a
detailed analysis of the field-dependent magnetization.
From exact-diagonalization calculations for extended Kitaev-Heisenberg Hamiltonians we 
additionally find that a transition from zigzag order to a spin-liquid ground state can be induced
in \aRuCl \ with external magnetic field.
\end{abstract}


\maketitle

\section{Introduction}
Quantum spin liquids (SL's) are states of matter that cannot be described by the broken symmetries
associated with conventional magnetic ground states \cite{balents10}.  Whereas there is a rich
variety of mathematical models that exhibit SL behavior, finding materials in which a quantum SL
state is realized is an intensely pursued goal in present day experimental condensed matter physics
\cite{Shimizu03,yamashita10,han12}.
Of particular interest is the Kitaev Hamiltonian on the
honeycomb lattice \cite{Kitaev06}, which is a mathematically well-understood two-dimensional model
exhibiting various topological SL states.  Its remarkable predicted properties include protection of
quantum information and the emergence of Majorana fermions \cite{Kitaev06,Pachos12}.

The search to realize the Kitaev model of effectively spin-1/2 particles on a honeycomb lattice
was centered until recently mainly on honeycomb iridate materials \cite{jackeli09,chaloupka10}
of the type $A_2$IrO$_3$, where $A$ is either Na or Li. 
In these systems though long-range magnetic order develops at low temperatures, for all known
different crystallographic phases \cite{Singh10,Ye12,choi12,Takayama14,Modic14}.
The SL regime is most likely preempted in the iridates by the presence of significant residual
Heisenberg-type $J$ couplings, by longer-range spin interactions, or by having crystallographically
distinct Ir-Ir bonds with dominant $J$'s on some of those, if not a combination of these factors
\cite{Kimchi11,rau14,Katukuri14,Nishimoto16}.

Also of interest in this context is ruthenium trichloride, RuCl$_3$, in its honeycomb crystalline
phase \cite{plumb14,majumder15,kubota15,sears15,sandilands15,banerjee16,kim15,Rousochatzakis15,Sandilands16}.
Very recent neutron scattering measurements suggest that the $4d^5$ halide honeycomb system is
closer to the Kitaev limit \cite{sandilands15,banerjee16}.
But also this material orders antiferromagnetically at low temperatures, as the $5d^5$ iridium
oxides do, and precisely how close to the idealized Kitaev model \aRuCl \ is, remains a question 
to be clarified.

Here we present results of combined quantum chemistry electronic-structure computations and
exact-diagonalization (ED) calculations for extended Kitaev-Heisenberg spin Hamiltonians using
as starting point for the ED study the magnetic couplings derived at the quantum chemistry
level.
Our results for the Ru$^{3+}$ $4d$-shell electronic structure show sizable trigonal splitting
of the $4d$ $t_{2g}$ levels and therefore a spin-orbit ground state that significantly deviates 
from the $j_{\rm eff}$=1/2 picture \cite{jackeli09}.
The trigonally distorted environment further gives rise to strong anisotropy of the computed $g$
factors, consistent with experimental observations \cite{kubota15,Miler68}.
Calculating the magnetic interactions between two adjacent $j_{\rm eff}$=1/2 moments, we find
that the nearest-neighbor (NN) Kitaev exchange $K$ is ferromagnetic (FM), in any of the \aRuCl \ 
crystalline structures reported recently.
It is however significantly weaker than in $5d^5$ Ir oxides and even than in 4$d^5$ Li$_2$RhO$_3$,
which points at a rather different balance between the various superexchange processes in the
halide and in the oxides.

The resulting magnetic phase diagram that we compute as function of longer-range second- and
third-neighbor magnetic couplings is very rich, due to the comparable size of the various
residual interactions.
While a SL state does show up in this phase diagram, it arises in a setting different from Kitaev's
original SL regime, as it emerges from an interplay of Kitaev physics and geometrically frustrated
magnetism.
We additionally find that applying an external magnetic field whilst the system is in the long-range 
ordered zigzag ground state can induce a phase transition into a quantum SL.
In order to make direct contact with experimental observations, we calculate by ED the field-dependent
magnetization in the presence of longer-range magnetic interactions and compare that to the
measurements.
This comparison makes clear that the ED and experimental data can only be matched when $J$ is small
and antiferromagnetic (AF) and $K$ significantly stronger and FM, in accordance with the results
from the {\it ab initio} quantum chemistry calculations. 

The magnitude of our computed $|K|$ compares well with recent estimates based on neutron scattering
data \cite{banerjee16}.
However, our finding that $K$ is FM brings into question the interpretation of the neutron scattering
experiments in terms of a pure Kitaev-Heisenberg model with AF $K$ but without longer-range magnetic
couplings which we find to be essential for an understanding of the magnetic properties of \aRuCl .

\section{Spin-orbit ground state and excitations}   
We start our discussion with the analysis of the Ru$^{3+}$ $4d$-shell electronic structure.
As in the $5d^5$ iridates, the magnetic moments in \aRuCl\ arise from the spin-orbit coupling
(SOC) of one hole in the transition-metal $t_{2g}$ subshell, described by the effective $L$=1
angular-momentum and $S$=1/2 spin quantum numbers.
Even if the SOC for $4d$ electrons is weaker than in the Ir $5d$ orbitals, it still splits the 
$t_{2g}^5$ states into a $j_{\rm eff}$=1/2 sector, where the hole resides, and a $j_{\rm eff}$=3/2
manifold that is filled.
For noncubic environment, these $j_{\rm eff}$=1/2 and $j_{\rm eff}$=3/2 components may display
some degree of admixture.

Three different crystallographic structures \cite{p312str,johnson15,Nagler16} have been 
reported for \aRuCl , each of those displaying finite amount of trigonal compression of the
Cl$_6$ octahedra.
To shed light on the nature of the 1/2-pseudospin in \aRuCl\ we first discuss in this section
results of {\it ab initio} many-body calculations at the complete-active-space self-consistent-field
(CASSCF) and multireference configuration-interaction (MRCI) levels of theory \cite{Helgaker2000}
for embedded atomic clusters having one RuCl$_6$ octahedron as reference unit.

As shown in Table\,\ref{t2g_dd}, the degeneracy of the Ru $t_{2g}$ levels is completely removed, with CASSCF
splittings of 69 and 72 meV when using the RuCl$_3$ $C2/m$ structure determined by Cao {\it et al.}
\cite{Nagler16}, a minimal active orbital space of only three 4$d$ orbitals, and no SOC. 
A ``trigonal'' orbital basis is used in Table\,\ref{t2g_dd} to express the $t_{2g}^5$ wave functions
\cite{Miler68}, in contrast to the Cartesian orbital basis employed for the Rh$^{4+}$ $t_{2g}^5$
states in Ref.\,\onlinecite{Katukuri15}, better suited for Li$_2$RhO$_3$ due to additional distortions 
of the ligand cages giving rise in the rhodate to one set of longer ligand-metal-ligand links
with an angle of nearly 180$^{\circ}$.

The corrections brought by the MRCI treatment are tiny, smaller than in the 4$d$ oxide \cite{Katukuri15}
Li$_2$RhO$_3$ due to less metal-$d$\,--\,ligand-$p$ covalency in the halide.
The smaller effective ionic charge at the ligand sites in the halide --- Cl$^-$ in RuCl$_3$ vs 
O$^{2-}$ in Li$_2$RhO$_3$, in a fully ionic picture --- further makes that the transition-metal 
$t_{2g}$--$e_g$ ligand-field splitting is substantially reduced in RuCl$_3$:
by MRCI calculations without SOC but with all five Ru $4d$ orbitals active in the reference
CASSCF, we find that the lowest $t_{2g}^4e_g^1$ ($t_{2g}^3e_g^2$) states are at only 1.3 (1.5)
eV above the low-lying $t_{2g}^5$ component (see Table\,\ref{dd_5.16}).
Interestingly, for the ``older'' $P3_112$ crystal structure proposed in Ref.\,[\onlinecite{p312str}],
we see that the $t_{2g}^3e_g^2$ sextet lies even below the lowest $t_{2g}^4e_g^1$ states, see 
Appendix A.
The smaller effective ligand charge might also be the cause for the smaller $t_{2g}$-shell
splittings in the halide: $\sim$70 meV in RuCl$_3$ (see caption of Table\,\ref{t2g_dd} and Table\,\ref{dd_5.16}) vs
$\sim$90 meV in Li$_2$RhO$_3$ \cite{Katukuri15} at the MRCI level, in spite of having similar
degree of trigonal compression in these two materials.

\begin{table}[!t]
\caption{
Ru$^{3+}$ $t_{2g}^5$ wave functions (hole picture) and relative energies (meV);
CASSCF results sans and with SOC for the crystal structure of Ref.\,[\onlinecite{Nagler16}].
Only the 4$d$ $t_{2g}$ orbitals were active in CASSCF;
by subsequent MRCI, the energies change to 0, 66, 73 sans SOC and to 0, 162, 201 
with SOC included.
Only one component of the Kramers' doublet is shown for each CASSCF+SOC relative energy.
$|\alpha\rangle$ corresponds to the $a_{1g}$ function while
$|\beta \rangle,\,|\gamma\rangle$ are $e'_g$ components \cite{Miler68}.
}
\begin{tabular}{llr}
\hline
\hline\\[-0.40cm]
$t_{2g}^5$ states &Relative  &Wave-function composition\\
(CASSCF)          &energies  &(normalized weights, \%)\\
\hline\\[-0.25cm]
Sans SOC\,:\\
$|\phi_1\rangle$  &0          &$99.75\,|\alpha \rangle + 0.25\,|\beta \rangle $\\
$|\phi_2\rangle$  &69         &$100\,|\gamma \rangle $\\
$|\phi_3\rangle$  &72         &$0.25\,|\alpha \rangle + 99.75\,|\beta \rangle $\\[0.10cm]

With SOC\,:\\[-0.02cm]
$|\psi_1\rangle$  &0         &$55\,|\phi_1,\downarrow\rangle + 23\,|\phi_2,\uparrow\rangle + 22\,|\phi_3,\uparrow\rangle$\\
$|\psi_2\rangle$  &157       &$45\,|\phi_1,\uparrow\rangle + 29\,|\phi_2,\downarrow\rangle + 26\,|\phi_3,\downarrow\rangle$\\
$|\psi_3\rangle$  &198       &                       $48\,|\phi_2,\uparrow\rangle + 52\,|\phi_3,\uparrow\rangle$\\
\hline
\hline
\end{tabular}
\label{t2g_dd}
\end{table}

With regard to the split $j_{\rm eff}$=3/2-like states that we compute at 195 and 234 meV by
MRCI+SOC calculations involving all three $t_{2g}^5$, $t_{2g}^4e_g^1$, and $t_{2g}^3e_g^2$
configurations in the spin-orbit treatment (see Table\,\ref{dd_5.16}),
clear excitations have been measured in that energy range in Raman scattering experiments
with ``crossed'' polarization geometries \cite{Sandilands16} and also in the optical
response of \aRuCl \ \cite{plumb14,Sandilands16}.
The peak observed at 140--150 meV by Raman scattering \cite{Sandilands16}, in particular, 
may find correspondence in the lowest $j_{\rm eff}$=3/2-like component that we compute at 195
meV.
It is interesting that in Sr$_2$IrO$_4$ the situation seems reversed as there the Raman
selection rules appear to favor the higher-energy split-off 3/2 states\cite{Yang15},
which are however shifted to somewhat lower energy as compared to resonant inelastic
x-ray scattering (RIXS)~\cite{Kim14}.
One should note however that in Sr$_2$IrO$_4$ the crystal-field physics is rather subtle, as
the local tetragonal distortion giving rise to elongated apical bonds is counteracted by
interlayer cation charge imbalance effects \cite{Bogdanov15}.

The rather broad feature at 310 meV in the imaginary part of the dielectric function has
been assigned to Ru$^{3+}$ $t_{2g}$-to-$e_g$ transitions \cite{Sandilands16}.
Our {\it ab initio} data do not support this interpretation, since the lowest
$t_{2g}$\,$\rightarrow$\,$e_g$ excitations are computed at $\approx$1.3 eV, but rather 
favor a picture in which the 310 meV peak corresponds to the upper 3/2-like component.
The latter can become optically active through electron-phonon coupling.
The rather large width of that excitation has been indeed attributed to electron-phonon
interactions in Ref.\,\onlinecite{Sandilands16}.

Comparison between our quantum chemistry results and the optical spectra \cite{plumb14,Sandilands16}
further shows that the experimental features at 1.2 and 2 eV, assigned in Ref.\,\onlinecite{Sandilands16}
to intersite $d$--$d$ transitions, might very well imply {\it on-site} Ru 4$d$-shell
excitations.
In particular, we find spin-orbit states of essentially $t_{2g}^4e_g^1$ nature at
1.3--1.5 eV and of both $t_{2g}^4e_g^1$ and $t_{2g}^3e_g^2$ character at 1.7--2.2 eV
relative energy, see Table\,\ref{dd_5.16}.
Experimentally the situation can be clarified by direct RIXS measurements on \aRuCl ,
for instance at the Ru $M_3$ edge.

\begin{table}[t]
\caption{
Ru$^{3+}$ $t_{2g}^me_g^n$ splittings (eV), with all five $4d$ orbitals active in CASSCF.
Except lowest line, each spin-orbit relative-energy entry implies a Kramers doublet.
Just the lowest and highest components are depicted for each group of $t_{2g}^4e_g^1$ 
spin-orbit states.
Only the $T$ and $A$ states shown in the table entered the spin-orbit calculations.
}
\begin{ruledtabular}
\begin{tabular}{lllll}
Ru$^{3+}$ $4d^5$          &CASSCF  &CASSCF           &MRCI   &MRCI\\
splittings                &        &+SOC             &       &+SOC\\
\hline
\\[-0.30cm]
$^2T_2$ ($t_{2g}^5$)      &0       &0                &0      &0   \\
                          &0.066   &0.193             &0.067  &0.195\\
                          &0.069   &0.232             &0.071  &0.234\\[0.22cm]
$^4T_1$ ($t_{2g}^4e_g^1$) &1.08    &1.25             &1.28   &1.33\\
                          &1.12    &$|$              &1.30   &$|$ \\
                          &1.13    &1.37             &1.31   &1.48\\[0.22cm]
$^4T_2$ ($t_{2g}^4e_g^1$) &1.76    &1.90             &1.97   &2.09\\
                          &1.81    &$|$              &2.01   &$|$ \\
                          &1.83    &1.98             &2.03   &2.17\\[0.22cm]
$^6\!A_1$ ($t_{2g}^3e_g^2$) &1.01    &1.09\,($\times 6$) &1.51   &1.74\,($\times 6$)\\
\end{tabular}
\end{ruledtabular}
\label{dd_5.16}
\end{table}

We have also calculated the magnetic $g$ factors in this framework.
By spin-orbit MRCI calculations with all five Ru $4d$ orbitals in the reference CASSCF,
we obtain for the $C2/m$ structure of Ref.\,[\onlinecite{Nagler16}]
$g_{xx}\!=\!g_{yy}\!=\!2.51$ and $g_{zz}\!=\!1.09$.
On the experimental side, conflicting results are reported for $g_{zz}$:
while Majumder {\it et al.}\cite{majumder15} derive from magnetic susceptibility data
$g_{zz}\!\sim\!2$, Kubota {\it et al.}\cite{kubota15} estimate a value $g_{zz}\!=\!0.4$.
The latter $g_{zz}$ value implies a rather large $t_{2g}$-shell splitting $\delta$, with 
$\delta/\lambda\!>\!0.75$ (see the analysis in Ref.\,\onlinecite{kubota15}).
The quantum chemistry $g$ factors are consistent with a ratio $\delta/\lambda\!\sim\!0.5$,
i.e., $t_{2g}$ splittings of $\approx$70 meV (see the data in Tables \ref{t2g_dd} and \ref{dd_5.16}) for a $4d$ SOC
in the range of 120--150 meV \cite{Miler68,kim15,Katukuri14c}.
Electron spin resonance measurements of $g$-factors might provide more detailed experimental information
that can be directly compared to our calculations.

\section{Intersite exchange for $j\!\approx\!1/2$ moments} 
NN exchange coupling constants were derived from MRCI+SOC calculations for embedded fragments
having two edge-sharing RuCl$_6$ octahedra in the active region.
As described in earlier work~\cite{Katukuri14,Katukuri15,Nishimoto16,Bogdanov15}, 
the {\it ab initio} data for the lowest four spin-orbit states describing the magnetic spectrum
of two NN octahedra is mapped in our scheme onto an effective spin Hamiltonian including both
isotropic Heisenberg exchange and symmetric anisotropic interactions.
Yet the spin-orbit calculations, CASSCF or MRCI, incorporate all nine triplet and nine singlet
low-energy states of predominant $t_{2g}^5$--$t_{2g}^5$ character.
As in earlier studies~\cite{Katukuri14,Katukuri15,Nishimoto16,Bogdanov15}, we account in the
MRCI treatment for all single and double excitations out of the valence $d$-metal $t_{2g}$ and
bridging-ligand $p$ shells.

For on-site Kramers-doublet states, the effective spin Hamiltonian for a pair of NN ions at
sites $i$ and $j$ reads
\begin{eqnarray}
{\cal H}_{i,j} =
           & &J\, \tilde{\bf{S}}_i \cdot \tilde{\bf{S}}_j
           +K \tilde{S}^z_i \tilde{S}^z_j
           +\sum_{\alpha \neq \beta} \Gamma_{\!\alpha\beta}(\tilde{S}^\alpha_i\tilde{S}^\beta_j +
                                                          \tilde{S}^\beta_i \tilde{S}^\alpha_j), \ \
\label{Eq:ham}
\end{eqnarray}
where $\tilde{\bf S}_i$ and $\tilde{\bf S}_j$ are 1/2-pseudospin operators, $J$ is the isotropic
Heisenberg interaction, $K$ the Kitaev coupling, and the $ \Gamma_{\alpha\beta}$ coefficients are
off-diagonal elements of the symmetric anisotropic exchange matrix with $\alpha,\beta\!\in\!\{x,y,z\}$.
Since the point-group symmetry of the Ru--Ru links is $C_{2h}$ in the $C/2m$ unit cell, the
antisymmetric Dzyaloshinskii-Moriya exchange is 0.
Also, $\Gamma_{zx}$=--$\Gamma_{yz}$ for $C_{2h}$ bond symmetry.
A local reference frame is used here, related to a given Ru-Ru link, as also employed in Refs.
\onlinecite{Katukuri14,Katukuri15,Nishimoto16}.
Details of the mapping procedure, {\it ab initio} data to effective spin Hamiltonian, are
described in Ref.\,\onlinecite{Bogdanov15} and Appendix A.

\begin{table}[!b]
\caption{
MRCI NN magnetic couplings (meV) for three different crystal structures proposed
for \aRuCl .
For the structure determined in Ref.\,[\onlinecite{johnson15}], the two crystallographically 
different NN Ru-Ru links are also different magnetically.
}
\begin{ruledtabular}
\begin{tabular}{|c|c| r c r c |}
Structure     &$\angle$Ru-Cl-Ru  &$K$ \   &$J$     &$\Gamma_{xy}$  &$\Gamma_{zx}$=--$\Gamma_{yz}$\\
\hline
$C2/m$ [\onlinecite{Nagler16}]
              &94$^{\circ}$      &$-5.6$   &$1.2$  &$-1.2$         &$-0.7$ \\
\hline
$C2/m$ [\onlinecite{johnson15}]  & & & &  & \\
Link\,1 ($\times\!2$)
              &94$^{\circ}$      &$-5.3$   &$1.2$  &$-1.1$         &$-0.7$ \\
Link\,2 ($\times\!1$)
              &93$^{\circ}$      &$-4.8$  &$-0.3$  &$-1.5$         &$-0.7$ \\
\hline
$P3_112$ [\onlinecite{p312str}]
              &89$^{\circ}$      &$-1.2$  &$-0.5$  &$-1.0$         &$-0.4$ \\
\end{tabular}
\end{ruledtabular}
\label{couplings}
\end{table}

From the quantum chemistry calculations, we obtain a FM Kitaev coupling $K$, for all three
crystalline structures recently reported in the literature (see Table\,\ref{couplings}).
Its strength is reduced as compared to the $4d^5$ honeycomb oxide Li$_2$RhO$_3$ \cite{Katukuri15},
with a maximum absolute value of 5.6 meV in the $C2/m$ structure proposed by Cao
{\it et al.}\,\cite{Nagler16}. 
We shall discuss and compare our finding of a FM Kitaev coupling to other theoretical and
experimental findings in the next section.
Anisotropic interactions of similar size, i.e., both $K$ and the off-diagonal couplings
$\Gamma_{\!\alpha\beta}$, are computed for the $C2/m$ configuration of Ref.\,[\onlinecite{johnson15}],
characterized by bond lengths and bond angles rather close to the values derived by Cao {\it et al.}
\cite{Nagler16}.
The Heisenberg $J$, on the other hand, changes sign with decreasing Ru-Cl-Ru angle but for the
bond angles reported in Refs.\,[\onlinecite{p312str},\onlinecite{johnson15},\onlinecite{Nagler16}]
and explicitly given in Table\,\ref{couplings} remains in absolute value smaller than $K$.

\begin{figure}[b]
\centering
\includegraphics[angle=270,width=0.98\linewidth]{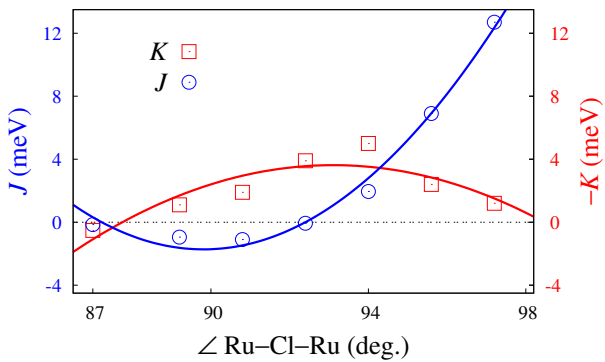}
\caption{
Variation of the NN Heisenberg and Kitaev couplings with the Ru-Cl-Ru angle in
model $C2/m$-type structures; results of spin-orbit MRCI calculations.
The NN Ru-Ru distance is set to 3.44 \AA \ and the Ru-Cl bond lengths are in each case
all the same.
The variation of the Ru-Cl-Ru angle is the result of gradual trigonal compression.
Curves are drawn just as a guide for the eye.
}
\label{fig:J_K}
\end{figure}

The trends we find with changing the Ru-Cl-Ru bond angle, apparent from Table\,\ref{couplings}, and earlier
results for the dependence of $K$ and $J$ on bond angles in oxide honeycomb compounds 
\cite{Katukuri15,Nishimoto16} motivate a more detailed investigation over a broader range of
Ru-Cl-Ru flexure.
The outcome of these additional calculations is illustrated in Fig.\,\ref{fig:J_K}.
In contrast to the oxides, where $|K|$ values in the range of 15--30 meV are computed for angles
of 98--100$^{\circ}$, the Kitaev coupling is never as large in RuCl$_3$.
$|K|$ shows a maximum of only $\approx$5 meV at 94$^{\circ}$ in Fig.\,1 and its angle dependence
is far from the nearly linear behavior in $4d^5$ and $5d^5$ oxides \cite{Katukuri15,Nishimoto16}. 

The Heisenberg $J$, on the other hand, displays a steep upsurge with increasing angle, more
pronounced as compared to the honeycomb oxides.
In other words, for large angles $J$ dominates in RuCl$_3$, in contrast to the results found in
$4d^5$ and $5d^5$ honeycomb oxides in the absence of bridging-ligand displacements parallel to
the metal-metal axis~\cite{Katukuri15,Nishimoto16}.
These notable differences between the halide and the oxides suggest a somewhat different balance
between the various superexchange processes in the two types of systems.

\section{Magnetic phase diagram}

\begin{figure}[t]
\centering
\includegraphics[width=0.9\linewidth]{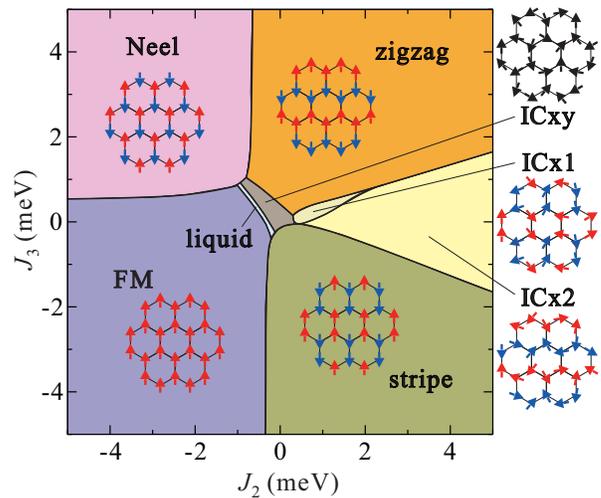}
\caption{
Phase diagram for the effective spin model of (\ref{Eq:ham}) supplemented by second- and
third-neighbor Heisenberg couplings $J_2$ and $J_3$.
MRCI NN interactions as listed on first entry in Table\,\ref{couplings} were used:
$J$=1.2, $K$=--5.6, $\Gamma_{xy}$=--1.2, $\Gamma_{zx}$=$-\Gamma_{yz}$=--0.7 (meV).
Schematic spin configurations for each particular phase are also shown.
No external field is applied in this set of calculations ($H$=0).
}
\label{fig:ed-1}
\end{figure}

To assess the consistency of our set of {\it ab initio} NN effective couplings with experimental
observations, we carried out ED calculations for the $\tilde S$=1/2 honeycomb model described
by (\ref{Eq:ham}) but including additionally the effect of second- and third-neighbor $J_2$ and
$J_3$ isotropic exchange.
Anisotropic longer-range interactions are however neglected since recent phenomenological
investigations conclude those are not sizable \cite{Winter16}.
We first considered the case without external magnetic field, $H\!=\!0$, and clusters of
24 $\tilde S$=1/2 sites with periodic boundary conditions (PBC's) as in previous studies
\cite{chaloupka10,Katukuri14}.
The static spin-structure factor
$S({\bf Q})\!=\!\sum_{ij}\langle ({\tilde{\bf S}}_i\!-\!\langle{\tilde{\bf S}}_i\rangle)\cdot
                                 ({\tilde{\bf S}}_j\!-\!\langle{\tilde{\bf S}}_j\rangle)\rangle
              \exp[i{\bf Q}\!\cdot\!({\bf r}_i\!-\!{\bf r}_j)]$
was calculated as function of variable $J_2$ and $J_3$ parameters while fixing the NN couplings
to the MRCI results computed for the crystalline structure of Ref.\,[\onlinecite{Nagler16}]
and listed in Table\,\ref{couplings}.

For a given set of $J_2$ and $J_3$ values, the dominant order is determined according to the
propagation vector ${\bf Q}\!=\!{\bf Q}_{max}$ providing a maximum value of $S({\bf Q})$.
As shown in Fig.\,\ref{fig:ed-1}, the phase diagram contains seven different phases: four
commensurate phases (FM, N\'{e}el, zigzag, stripy), three with incommensurate (IC) order
(labeled as ICx1, ICx2, ICxy), and a SL phase.
The ICx1 and ICx2 configurations have the same periodicities along the $b$ direction as the
stripy and zigzag states, respectively, and display an IC wave number along $a$.
The ICxy phase has IC propagation vectors along both $a$ and $b$.
The variety of IC phases in the computed phase diagram is related to the comparable strength
of the NN $J$ and the off-diagonal NN couplings $\Gamma_{\alpha\beta}$.
For example, the system is in the ICxy state at $J_2\!=\!J_3\!=\!0$.
From the experimental observations, the low-temperature magnetic structure of \aRuCl \ is
$ab$-plane zigzag AF order~\cite{sears15,johnson15,Nagler16}.
We find that the zigzag state is stabilized in a wide range of AF $J_2$ and $J_3$ values in 
our phase diagram. 

\begin{figure}[b]
\centering
\includegraphics[width=0.9\linewidth]{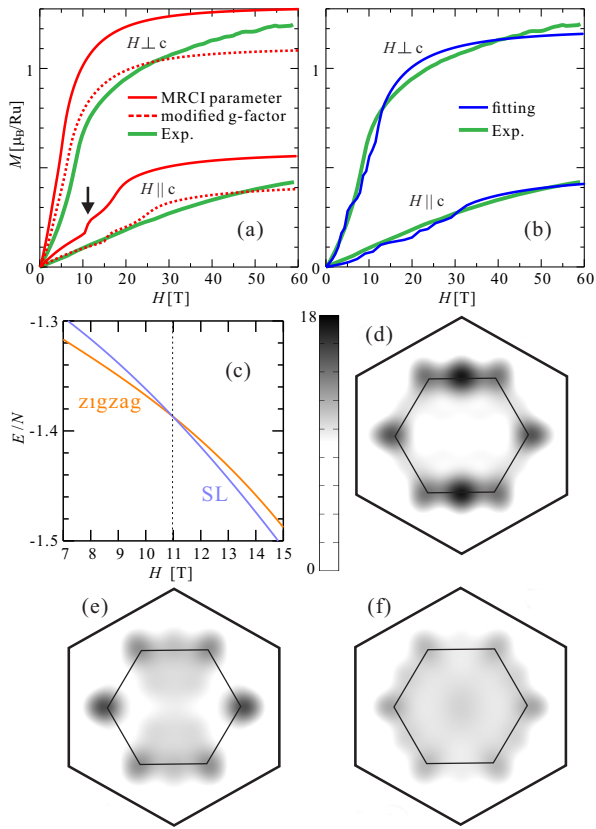}
\caption{
(a) Comparison between the measured magnetization curves \cite{johnson15} and ED results using
MRCI $g$ factors and NN couplings plus $J_2\!=\!J_3\!=\!0.25$ meV.
The dashed lines show ED results with modified $g$ factors, $g_{xx}\!=\!g_{yy}\!=\!2.30$, 
$g_{zz}\!=\!0.83$.
(b) ED-based fit of the magnetization curves with no constraints on the NN interactions.
(c) Energies of the lowest two magnetic states (zigzag and SL) around the level-crossing point
$H\!=\!10.8$ T (${\bf H}\!\parallel\!c$).
Static spin-structure factors $S({\bf Q})$ are shown for
(d) $H\!=\!0$,
(e) $H\!=\!10.4$,
(f) $H\!=\!11.2$ T.
}
\label{fig:ed-2}
\end{figure}

To estimate the strength of $J_2$ and $J_3$ in \aRuCl , we performed a fitting of the experimental
magnetization curves~\cite{johnson15} using data from the ED calculations.
Different signs for $J$ and $K$ determine qualitatively different shapes for the magnetization
curves.
We find that $J\!>\!0$ and $K\!<\!0$ values are required to reproduce the overall pattern
of the measured magnetization, which exhibits a very slow saturation with increasing external
field (see Appendix B).
Additionally, AF values for $J_2$ and $J_3$ significantly shift the saturation to higher field
and therefore these longer-range couplings must be small ($\lesssim$1 meV, see Appendix B) to reproduce
the experiment.
The observed magnetization curves are compared to ED results in Fig.~\ref{fig:ed-2}(a), for
${\bf H}\!\perp\!c$ and ${\bf H}\!\parallel\!c$.
We used MRCI $g$ factors and NN interactions and set $J_2\!=\!J_3\!=\!0.25$ meV in a first set
of ED calculations.
Despite some quantitative deviations, the overall shapes of the experimental curves are
reproduced in the calculations.

The magnetization is in fact very sensitive to the $g$ factors.
Quite reasonable fits can be then obtained by rather small modification of these quantities,
e.\,g., from the MRCI values $g_{xx}\!=\!g_{yy}\!=\!2.51$ and $g_{zz}\!=\!1.09$ to
                             $g_{xx}\!=\!g_{yy}\!=\!2.30$ and $g_{zz}\!=\!0.83$.
Even better agreement with the experiment is finally achieved by removing any constraint for
the NN effective exchange couplings.
A best fit is found with $J\!=\!2.0$, $K\!=\!-10.0$, $J_2\!=\!J_3\!=\!0.5$, $g_{xx}\!=\!g_{yy}\!=\!2.4$,
and $g_{zz}\!=\!0.95$ in the ED calculations for the extended $J$-$K$-$J_2$-$J_3$ model
[see Fig.~\ref{fig:ed-2}(b)].
These ``fitted'' $J$ and $K$ NN interactions are comparable to those derived by MRCI.
It is however difficult to extract information on the NN off-diagonal exchange terms by using
fits to the experimental data because the magnetization is not very sensitive to these
off-diagonal couplings.

Most interestingly, a level crossing between the lowest two states (zigzag and SL) is seen at
$H\!=\!10.8$ T for ${\bf H}\!\parallel\!c$.
The static spin-structure factors for $H$=0, 10.4, and 11.2 T are plotted in
Fig.~\ref{fig:ed-2}(d)-(f).
The zigzag AF order is gradually weakened with increasing $H$, destroyed at $H\!=\!10.8$ T, 
and instead a SL ground state occurs for $H\!>\!10.8$ T.
The SL state can be confirmed by a featureless static spin-structure factor.
There are in principle two possible origins of the SL state:
one is the relatively large Kitaev interaction and the other is the frustration of exchange
interactions beyond the Kitaev model.
Only the NN spin-spin correlations being large at $H\!=\!10.8\!-\!14.2$ T (see Appendix B for
details on the spin-spin correlation functions) is indicative of a Kitaev-like SL regime.
However, the MRCI calculations indicate $|K|/J$ ratios in the range of 3--5 for the $C/2m$
structures (see Table\,\ref{couplings}) while a commonly used criterion \cite{chaloupka10} for
identifying the Kitaev SL is having $|K|/J\!>\!7.8$, so that the further frustration of magnetic
interactions is relevant as well.
One simple way to rationalize these findings is that an external field effectively weakens the
effect of the AF NN $J$ due to partial spin polarization and consequently $|K|/J$ is effectively
enhanced.
Another way of qualitatively appreciating this point is that when one looks at the $J_2$-$J_3$
phase diagram in Fig.~\ref{fig:ed-1}, the main features of which are very similar to those
\cite{Katukuri14} found for Na$_2$IrO$_3$, a trajectory in the phase diagram from zigzag order
(the low field state) to a saturated ferromagnet (the very high field state) is likely to pass
through the SL phase.
It is interesting that such a field-induced SL state due to frustration has been also predicted recently for the $S$=1/2 AF
kagom\'{e} lattice~\cite{nishimoto13}.

\section{Discussion}
Our finding of a FM Kitaev interaction can be first compared with the conclusions of other
theoretical investigations.
In fact, the analysis of effective superexchange models using hopping matrix elements and 
effective Hubbard-$U$ interactions derived from density-functional (DF) electronic-structure
calculations lead to contradictory results:
an AF NN Kitaev coupling has been earlier predicted by Kim {\it et al.} \cite{kim15} and a FM
$K$ has been more recently found by Winter {\it et al.} \cite{Winter16}.
Our result is consistent with the latter.
Relevant in this regard is further the trends we observe for the effective $K$ by running
spin-orbit calculations at different levels of approximation: restricted open-shell Hartree-Fock 
(ROHF), CASSCF, and MRCI.
The respective $K$ values are 1.2, --2.5, and --5.6 meV, for the $C2/m$ structure of
Ref.\,[\onlinecite{Nagler16}].
It is seen that accounting for intersite $t_{2g}$--$t_{2g}$ hopping by CASSCF changes the sign of
$K$ from AF to FM and that by additionally taking into account the bridging-ligand $3p$ and metal
$e_g$ levels by MRCI calculations with single and double excitations only pushes $K$ more on the
FM side.
It is unlikely that additional excitations, ``triple'' etc., would change the sign of $K$ back
to the AF ROHF.

To make direct contact with experimental observations, one can compare the measured field-dependent
magnetization with the theoretical results, as we did above, finding that only $J\!>\!0$ and
$K\!<\!0$ are consistent with the measurements \cite{johnson15}.
This however contradicts the interpretation of recent inelastic neutron scattering data on the
magnetic excitation spectrum \cite{banerjee16}, according to which $K$ is very similar in
magnitude to our finding but AF.

This point remains to be clarified but a possible explanation is related to modeling the
experimental magnetic excitation spectra in the zigzag ordered state in terms of a pure
Kitaev-Heisenberg Hamiltonian without longer-range couplings.
In such a restricted model, zigzag order can {\it only} occur when $J\!<\!0$ and $K\!>\!0$ --- 
using the zigzag ordered ground state as input for the pure Kitaev-Heisenberg model fixes
$K\!>\!0$ from the beginning and
a description of the magnetic excitations on top of this ground state in terms of linear 
spin-wave theory is necessarily confined to this boundary condition.
We find however that \aRuCl\ is in a parameter regime where {\it without} longer-range,
second-neighbor and third-neighbor, interactions the ordering pattern would be an incommensurate
AF state (see Fig. 2) which is close to the stripe-like AF phase.
This is the consequence of $J\!>\!0$ and $K\!<\!0$.
A weak AF third-neighbor exchange $J_3$ is essential to stabilize the zigzag order that is
experimentally observed --- this zigzag ground state is driven by the geometric magnetic frustration
induced by $J_3$ and consistent with $K$ being dominant and FM.

For an interpretation of the magnon features in the neutron spectrum Ref.~\onlinecite{banerjee16}
relies on linear spin-wave theory while for resolving the signatures of the fractionalized
excitations --- the actual fingerprint of the system being proximate to a Kitaev SL state ---
relies on a comparison to a Kitaev-only Hamiltonian.
This should provide a full quantum description of the relevant physics on energy scales larger
than weak interlayer magnetic couplings.
The Kitaev point is particularly interesting because exact statements can be made
\cite{Kitaev06,knolle14,Nussinov13}.
In the honeycomb Kitaev model the excitations are exactly fractionalized into localized fluxes
and delocalized Majorana modes.
Its dynamic spin-structure factor, which determines the inelastic neutron scattering response,
is dominated by a spin excitation creating two fluxes.
As the fluxes are localized, the spin-structure factor is rather dispersionless and only a weak
momentum dependence arises from screening of the fluxes by gapless Majorana modes \cite{knolle14}.
The sign of $K$ sets the sign for the dispersion of these Majorana modes that screen the fluxes
\cite{Kitaev06}.
The upshot is that the dynamic structure factor in the Kitaev model strongly depends on the
magnitude of $|K|$ (which sets the energy threshold for flux creation) but only very weakly on
its sign --- fits to the data with $|K|$ and $-|K|$ then provide very similar results.

\section{Conclusions}
Quantum chemistry calculations show that in \aRuCl\ there is sizable trigonal splitting of
the Ru $4d^5$ levels.
This results in splitting of the spin-orbital excitation energies, which can be measured by
e.\,g. resonant inelastic x-ray scattering, and in admixture of the $j_{\mathrm{eff}}$=1/2
and $j_{\mathrm{eff}}$=3/2 states.
The resulting anisotropy of the magnetic $g$ factors that we compute is consistent with
experimental observations~\cite{kubota15,Miler68}.

The nearest-neighbor Heisenberg interaction $J$ is found to be weak and antiferromagnetic in
the {\it ab initio} computations while the Kitaev $K$ is 3--5 times larger and ferromagnetic.
Using these magnetic couplings as a basis for effective-model exact-diagonalization
calculations of the magnetic phase diagram, we show that $J>0$ and $K<0$ values are required
to reproduce the shape of the observed magnetization.
The latter exhibits a very slow saturation with increasing the external field.
As residual longer-range magnetic interactions would significantly shift the saturation to
higher field, they must be small.
At the same time, however, we find these longer-range couplings are essential in producing the
experimentally observed zigzag magnetic order in \aRuCl .

We also determine by quantum chemistry calculations the dependence of the NN $K$ and $J$ 
interactions on the angle defined by two adjacent metal sites and a bridging ligand.
Along with similar curves we compute for the ``213'' honeycomb compounds \cite{Katukuri15,Nishimoto16}
--- Na$_2$IrO$_3$, Li$_2$IrO$_3$, and Li$_2$RhO$_3$ --- these results provide theoretical
benchmarks for strain and pressure experiments on 4$d^5$/5$d^5$ honeycomb halides and oxides.

At $H$=10 T, a level crossing between the lowest two states is seen for field along the $c$
direction, inducing a transition from zigzag order to a spin-liquid state.
Our calculations suggest that not only \aRuCl\ but also Na$_2$IrO$_3$ is a candidate material
to observe such a transition, either at low-temperature ambient conditions or under external
pressure.

 \ \ \

{\it Acknowledgements.} \
We thank A.~Tsirlin and S.~E.~Nagler for fruitful discussions.
We also thank S.-H.\,Kim and B.\,B\"uchner for discussions and for sharing unpublished
experimental data.
S.\,N. and L.\,H. acknowledge financial support from the German Research Foundation 
(Deutsche Forschungsgemeinschaft, DFG --- SFB-1143 and HO-4427).
J.v.d.B. acknowledges support from the Harvard-MIT CUA.

\section{Appendix A\,:\, Quantum chemistry calculations}

{\bf Ru$^{3+}$ 4$d$-shell electronic structure}.\,  
{\it Ab initio} many-body quantum chemistry calculations were first carried out to establish the 
nature of the Ru$^{3+}$ 4$d^5$ ground state and lowest Ru 4$d$-shell excitations in RuCl$_3$.
An embedded cluster having as central region one [RuCl$_6$]$^{3-}$ octahedron was used.
To describe the finite charge distribution in the immediate neighborhood, the three adjacent
RuCl$_6$ octahedra were also explicitly included in the quantum chemistry computations while the
remaining part of the extended solid-state matrix was modeled as a finite array of point 
charges fitted to reproduce the ionic Madelung field in the cluster region \cite{Ewald_soft}.
Energy-consistent relativistic pseudopotentials were used for the central Ru ion, along with
valence basis sets of quadruple-zeta quality augmented with two $f$ polarization functions
\cite{Peterson07}.
For the Cl ligands of the central RuCl$_6$ octahedron, we employed all-electron valence
triple-zeta basis sets \cite{Woon93}.
For straightforward and transparent analysis of the on-site multiplet physics (see Table\,\ref{dd_5.16} in
main text and Table\,\ref{dd_p312} in this section), the adjacent Ru$^{3+}$ sites were described as closed-shell
Rh$^{3+}$ $t_{2g}^6$ ions, using relativistic pseudopotentials and valence triple-zeta basis
functions \cite{Peterson07}.
Ligands of these adjacent octahedra that are not shared with the central octahedron were modeled
with all-electron minimal atomic-natural-orbital basis sets \cite{Pierloot95}.
Results in excellent agreement with the experiment were found by using such a procedure in, e.\,g.,
Sr$_2$IrO$_4$ \cite{Bogdanov15} and CaIrO$_3$ \cite{Ir113_bogdanov_12}.

All computations were performed with the {\sc molpro} quantum chemistry package \cite{Molpro12}.
To access the Ru on-site excitations, we used active spaces of either three (see Table\,\ref{t2g_dd} in main
text) or five (Table\,\ref{dd_5.16} in main text and Table\,\ref{dd_p312} in this section) orbitals in CASSCF.
In the subsequent MRCI \cite{Werner88,Knowles92}, the Ru $t_{2g}$ and Cl 3$p$ electrons at the
central octahedron were correlated.
The Pipek-Mezey localization module \cite{Pipek89} available in {\sc molpro} was employed for
separating the metal 4$d$ and Cl 3$p$ valence orbitals into different groups, i.\,e.,
centered at sites of either the central octahedron or the adjacent octahedra.
The spin-orbit treatment was carried out as described in Ref.\,\cite{SOC_molpro}.

One important finding in our quantum chemistry investigation is that compared to the $4d$ and $5d$
oxide honeycomb systems --- Li$_2$RhO$_3$, Li$_2$IrO$_3$, Na$_2$IrO$_3$ --- the smaller ligand 
ionic charge in the halide gives rise to significantly weaker $t_{2g}$--$e_g$ splittings.
This is apparent in Table \ref{dd_5.16} in the main text:
for the $C2/m$ crystalline structure of Cao {\it et al.} \cite{Nagler16}, we compute excitation 
energies of only $\approx$1.3 eV for the lowest $t_{2g}^4e_g^1$ states.
Even more suggestive in this regard is the energy-level diagram we compute for the $P3_112$ 
crystalline structure of Ref.\,\cite{p312str}. 
For the latter, the sequence of Ru$^{3+}$ $t_{2g}^me_g^n$ levels is shown in Table \ref{dd_p312}:
it is seen that the $^6\!A_1$ ($t_{2g}^3e_g^2$) state is even lower in energy than $^4T_1$
($t_{2g}^4e_g^1$). 
Such low-lying $t_{2g}^me_g^n$ excited states may obviously play a more important role than in
the oxides in intersite superexchange.

\begin{table}[t]
\caption{
Ru$^{3+}$ $t_{2g}^me_g^n$ splittings (eV) in the crystalline structure of Ref.\,\cite{p312str}.
Except the $t_{2g}^3e_g^2$ states, each spin-orbit relative-energy entry implies a Kramers doublet.
Just the lowest and highest components are depicted for each group of $t_{2g}^4e_g^1$ 
spin-orbit states.
Only the $T$ and $A$ states shown in the table entered the spin-orbit calculations.
}
\begin{ruledtabular}
\begin{tabular}{lllll}
Ru$^{3+}$ 4$d^5$            &CASSCF &CASSCF             &MRCI   &MRCI\\
splittings                  &       &+SOC               &       &+SOC\\
\hline
\\[-0.20cm]
$^2T_2$ ($t_{2g}^5$)        &0      &0                  &0      &0   \\
                            &0.04   &0.16               &0.05   &0.19\\
                            &0.05   &0.16               &0.06   &0.23\\[0.22cm]
$^6\!A_1$ ($t_{2g}^3e_g^2$) &0.07   &0.21\,($\times 6$) &0.92   &0.92\,($\times 6$)\\[0.22cm]
$^4T_1$ ($t_{2g}^4e_g^1$)   &0.62   &0.78               &0.94   &1.10\\
                            &0.66   &$|$                &0.97   &$|$ \\
                            &0.66   &0.85               &0.98   &1.23\\[0.22cm]
$^4T_2$ ($t_{2g}^4e_g^1$)   &1.27   &1.42               &1.52   &1.65\\
                            &1.33   &$|$                &1.56   &$|$ \\
                            &1.38   &1.55               &1.63   &1.77\\
\end{tabular}
\end{ruledtabular}
\label{dd_p312}
\end{table}

\begin{table*}[t]
\caption{
Matrix elements of the {\em ab initio} model Hamiltonian (meV), as obtained by spin-orbit
MRCI.
The two-site singlet and (split) triplet states are labeled $|s \rangle$ and
\{$|t_{x}\rangle$,\,$|\tilde t_{y}\rangle$,\,$|\tilde t_{z}\rangle$\}, respectively.
$|\tilde t_{y}\rangle$ and $|\tilde t_{z}\rangle$ are admixtures of `pure' $|1,-1\rangle$
and $|1,0\rangle$ spin functions.
}
\begin{ruledtabular}
\begin{tabular}{rcccc}

$H^{kl}_\mathit{ab\,initio}$

                 &$|\tilde t_{y}\rangle$ &$|t_{x}\rangle$ &$|s\rangle$ &$|\tilde t_{z}\rangle$ \\[0.10cm]

$\langle\tilde t_{y}|$ &$0$              &$0.804i\,\mu_B H_y+2.720i\,\mu_B H_z$

                                                      &$0$

                                                                        &$-1.826i\,\mu_B H_x$ \\[0.10cm]

$\langle t_{x}|$

                 &$-0.804i\,\mu_B H_y-2.720i\,\mu_B H_z$

                                   &$1.189$         &$0$

                                                                        &$-0.1.130i\,\mu_B H_y-0.280i\,\mu_B H_z$ \\[0.10cm]

$\langle s|$

                 &$0$

                                   &$0$

                                                      &$2.187$        &$0$ \\[0.10cm]

$\langle\tilde t_{z}|$

                 &$1.826i\,\mu_B H_x$              &$0.1.130i\,\mu_B H_y+0.280i\,\mu_B H_z$

                                                      &$0$

                                                                        &$3.475$ \\

\end{tabular}
\end{ruledtabular}
\label{abinitioHam}
\end{table*}

\begin{table*}[!t]
\caption{
Matrix form of the effective spin Hamiltonian in the basis of zero-field eigenstates.
$\Gamma^{-}$ stands for $\Gamma_{yy}^{'}-\Gamma_{zz}^{'}$\,; expressions for the $\Delta$ and $\Omega$
terms are provided in text.} 
\begin{ruledtabular}
\begin{tabular}{rcccc}

$H_\mathrm{eff}^{kl}$

                 &$|\tilde t_{y}\rangle$ &$|t_{x}\rangle$ &$|s\rangle$ &$|\tilde t_{z}\rangle$ \\[0.10cm]

$\langle\tilde t_{y}|$ &$0$              &$iH_y \Delta_y +iH_z \Delta_z$

                                                      &$0$

                                                                        &$ig_{xx} H_x$ \\[0.10cm]

$\langle t_{x}|$

                 &$-iH_y \Delta_y -iH_z \Delta_z$

                                   &$\frac{1}{4} (3\Gamma^{-}+\sqrt{4\Gamma_{yz}^{'2}+(\Gamma^{-})^2}+6\Gamma_{zz}^{'})$         &$0$

                                                                        &$iH_y \Omega_y +iH_z \Omega_z$ \\[0.10cm]

$\langle s|$

                 &$0$

                                   &$0$

                                                      &$\frac{1}{4} (\Gamma^{-}+\sqrt{4\Gamma_{yz}^{'2}+(\Gamma^{-})^2}+2\Gamma_{zz}^{'}-4J^{'})$        &$0$ \\[0.10cm]

$\langle\tilde t_{z}|$

                 &$-ig_{xx} H_x$              &$-iH_y \Omega_y -iH_z \Omega_z$

                                                      &$0$

                                                                        &$\frac{1}{2} \sqrt{4\Gamma_{yz}^{'2}+(\Gamma^{-})^2}$ \\

\end{tabular}
\end{ruledtabular}
\label{modHam}
\end{table*}

Ru 4$d^5$ $g$ factors were computed following the procedure described in Ref.\,\cite{Bogdanov15}.
The values provided in the main text, $g_{xx}\!=\!g_{yy}\!=\!2.51$ and $g_{zz}\!=\!1.09$, were
obtained by including the $^2T_2$ ($t_{2g}^5$), $^4T_1$ ($t_{2g}^4e_g^1$), $^4T_2$ ($t_{2g}^4e_g^1$),
and $^6\!A_1$ ($t_{2g}^3e_g^2$) states in the spin-orbit treatment.
The orbitals were optimized for an average of all these states.
The $z$ axis is here taken along the trigonal axis, perpendicular to the honeycomb plane of Ru ions.
The strength of the coupling to external magnetic field can also be extracted from more
involved calculations as described in the next subsection.

{\bf Intersite exchange.}\, 
NN magnetic coupling constants were derived from CASSCF+MRCI spin-orbit calculations on
units of two edge-sharing [RuCl$_6$]$^{3-}$ octahedra.
Similar to the computations for the on-site excitations, the four octahedra adjacent to the
reference [Ru$_2$Cl$_{10}$]$^{4-}$ entity were also included in the actual (embedded)
cluster.
We used energy-consistent relativistic pseudopotentials along with valence basis sets of
quadruple-zeta quality for the two Ru cations in the reference unit \cite{Peterson07}.
All-electron basis sets of quintuple-zeta quality were employed for the bridging ligands and
triple-zeta basis functions for the remaining chlorine anions of the reference octahedra
\cite{Woon93}.
We further utilized two $f$ polarization functions \cite{Peterson07} for each Ru ion of the
central, reference unit and four $d$ polarization functions \cite{Woon93} at each of the
bridging ligand sites.
Ru$^{3+}$ ions of the four adjacent octahedra were modeled as closed-shell Rh$^{3+}$ species,
following a strategy similar to the calculations for the on-site $4d$-shell transitions.
The same computational scheme yields magnetic coupling constants in very good agreement with
experimental estimates in CaIrO$_3$ \cite{Ir113_bogdanov_12}, Ba$_2$IrO$_4$ \cite{Katukuri14b},
and Sr$_2$IrO$_4$ \cite{Bogdanov15,Ir214_lupascu_14}.

The mapping of the {\it ab initio} quantum chemistry data onto the effective spin model defined
by (\ref{Eq:ham}) implies the lowest four spin-orbit states associated with the different 
possible couplings of two NN 1/2 pseudospins.
The other 32 spin-orbit states within the $t_{2g}^5$--$t_{2g}^5$ manifold \cite{Katukuri14,Katukuri15}
involve $j_{\mathrm{eff}}\!\approx\!3/2$ to $j_{\mathrm{eff}}\!\approx\!1/2$ charge excitations
\cite{jackeli09,Katukuri15} and lie at $\gtrsim$150 meV higher energy (see Tables \ref{t2g_dd}, 
\ref{dd_5.16} and Refs.\,\cite{Katukuri15,Katukuri14c}), an energy scale much larger than the
strength of intersite exchange.
To derive numerical values for all effective spin interactions allowed by symmetry in (\ref{Eq:ham}),
we additionally consider the Zeeman coupling
\begin{equation}
\hat{\cal H}_{i,j}^Z=\sum_{q=i,j}\mu_B ({\bf L}_{q} + g_e {\bf S}_{q}) \cdot \bf{H} \,,
\end{equation}
where ${\mathbf L}_q$ and ${\mathbf S}_q$ are angular-momentum and spin operators at a given Ru site
while $g_e$ and $\mu_B$ stand for the free-electron Land\'{e} factor and Bohr magneton, respectively
(see also Ref.\,\onlinecite{Bogdanov15}).
Each of the resulting matrix elements $H^{kl}_\mathit{ab\,initio}$ computed at the quantum chemistry
level, see Table \ref{abinitioHam}, is assimilated to the corresponding matrix element
$H^{kl}_{\mathrm{eff}}$ of the effective spin Hamiltonian, see Table \ref{modHam}.
This one-to-one correspondence between {\it ab initio} and effective-model matrix elements
enable an assessment of all coupling constants in (\ref{Eq:ham}).

For $C_{2h}$ symmetry of the [Ru$_2$Cl$_{10}$] unit \cite{Nagler16}, it is convenient to choose
a reference frame with one of the axes along the Ru-Ru link.
The data collected in Tables \ref{abinitioHam} and \ref{modHam} is expressed by using such a
coordinate system, with the $x$ axis along the Ru-Ru segment and $z$ perpendicular to the
Ru$_2$Cl$_2$ plaquette.
The $\bar{\bar{\Gamma}}$ tensor reads then
\begin{equation}
 \bar{\bar{\Gamma}} =
 \left(
 \begin{array}{lll}
  \Gamma_{xx}^{'}&0           &\hspace{0.2cm} 0 \\ \vspace{0.1cm}
 0                        &\Gamma_{yy}^{'} &\hspace{0.2cm}  \Gamma_{yz}^{'} \\ \vspace{0.1cm}
 0                        &\Gamma_{yz}^{'}           &\hspace{0.2cm} \Gamma_{zz}^{'}
 \end{array}
 \right), 
\end{equation}
where $\Gamma_{xx}^{'}=-\Gamma_{yy}^{'}-\Gamma_{zz}^{'}$ and the ``prime'' notation refers to
this particular coordinate system.
The Kitaev-like reference frame within which the data in Table\,\ref{couplings} is expressed implies a
rotation by 45$^\circ$ about the $z$ axis \cite{Katukuri14,Katukuri15,Nishimoto16}.
The connection between the parameters of Table\,\ref{couplings}, corresponding to the Kitaev-like axes,
and the ``prime'' quantities in Tables \ref{abinitioHam} and \ref{modHam} is given by the
following relations \cite{Katukuri14,Katukuri15,Nishimoto16}\,:
\begin{equation}
\begin{aligned}
  J=J^{'} + \frac{\Gamma_{xx}^{'}+\Gamma_{yy}^{'}}{2}, \,\,\,\, K=-\frac{3(\Gamma_{xx}^{'}+\Gamma_{yy}^{'})}{2}\\
 \Gamma_{xy}=\frac{\Gamma_{xx}^{'}-\Gamma_{yy}^{'}}{2}, \,\,\,\, \Gamma_{zx}=-\Gamma_{yz}=-\frac{\Gamma_{yz}^{'}}{\sqrt{2}}.\\
\end{aligned}
\end{equation}
The terms $\Delta_n$ and $\Omega_n$ in Table \ref{modHam} (where $n\!\in\!\{y,z\}$) stand for\,:
{\footnotesize
\begin{equation}
  \Delta_n = \frac{2\Gamma_{yz}^{'}g_{yn}+(-\Gamma_{yy}^{'}+\Gamma_{zz}^{'}-\sqrt{4\Gamma_{yz}^{'2}+(\Gamma_{yy}^{'}-\Gamma_{zz}^{'})^2})g_{nz}}{\sqrt{4\Gamma_{yz}^{'2}+(\Gamma_{yy}^{'}-\Gamma_{zz}^{'}+\sqrt{4\Gamma_{yz}^{'2}+(\Gamma_{yy}^{'}-\Gamma_{zz}^{'})^2})^2}}\,,
\end{equation}
\begin{equation}
 \Omega_n = \frac{2\Gamma_{yz}^{'}g_{yn}+(-\Gamma_{yy}^{'}+\Gamma_{zz}^{'}+\sqrt{4\Gamma_{yz}^{'2}+(\Gamma_{yy}^{'}-\Gamma_{zz}^{'})^2})g_{nz}}{\sqrt{4\Gamma_{yz}^{'2}+(\Gamma_{yy}^{'}-\Gamma_{zz}^{'}-\sqrt{4\Gamma_{yz}^{'2}+(\Gamma_{yy}^{'}-\Gamma_{zz}^{'})^2})^2}}\,.
\end{equation}
}

\section{Appendix B\,:\, ED calculations}

\begin{figure}[!b]
\includegraphics[width=0.85\linewidth]{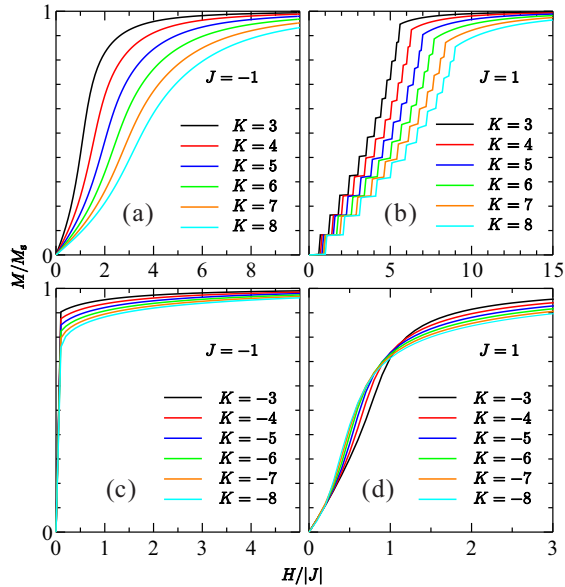}
\caption{
Magnetization curves of the pure Kitaev-Heisenberg model for
(a) $J<0$, $K>0$,
(b) $J>0$, $K>0$,
(c) $J<0$, $K<0$, and
(d) $J>0$, $K<0$.
The magnetic field is applied along the $z$ direction and the saturation of the magnetization is
set to be $M=M_s$.
}
\label{fig:S1}
\end{figure}

{\bf Magnetization curves for the Kitaev-Heisenberg model}.\,  
Magnetization curves of the pure Kitaev-Heisenberg model, calculated by ED on a 24-site
cluster, are plotted in Fig.~\ref{fig:S1}.
The overall shapes are qualitatively well determined once the signs of $J$ and $K$ are fixed.
For $J>0$ and $K>0$ [Fig.\,\ref{fig:S1}(b)],      
the magnetization increases linearly at low field and more steeply at higher
field.
This behavior is similar to that of the two-dimensional (2D) bipartite Heisenberg systems;
the main difference is the existence of a kink near the saturation, due to local AF interactions
and the mixing of different $S^z$-sectors.
Below the kink, the NN spin correlations remain AF.
%
%
For $J<0$ and $K<0$ [Fig.\,\ref{fig:S1}(c)], the magnetization jumps up to a finite value at $H=0^+$ and gradually saturates
with increasing field.
This gradual saturation is the result of local FM interactions $S^xS^x$ and $S^yS^y$.
%
%
For $J<0$ and $K>0$ [Fig.\,\ref{fig:S1}(a)], the magnetization increases linearly at low field, reflecting the AF $J$,
smoothly connects to the higher-field curve, and then saturates gradually with increasing field,
similar to the case of $J<0$ and $K<0$.
This qualitative behavior
is basically the result of competing FM $J$ and AF $K$.
When $K$ is small, the magnetization saturates rapidly with increasing field due to the FM $J$;
as the AF $K$ increases, the saturation is shifted to higher field.
The shape of the magnetization curve itself is almost unchanged with changing $K$ and the magnetic
field can be simply rescaled by $K\!\cdot\!H$.
Typically, the effect of FM $K$ on the magnetization curve is small but the saturation becomes
slower for larger $K$.
A linear increase in weak fields and very slow saturation at higher fields was experimentally
observed for \aRuCl.
Such behavior is found in the calculations only for $J>0$ and $K<0$ [Fig.\,\ref{fig:S1}(d)].

Generally, the magnetization curve of the Heisenberg model is a step function in calculations on
finite-size systems, due to discrete effects.
However, in the Kitaev-Heisenberg model, the total $S^z$ is no longer conserved due to terms such
as $S^+S^+$ and $S^-S^-$.
The magnetization curve can be then a smooth function.
In our results, small steps are still visible in the magnetization curve for the case of $J>0$ and
$K>0$.
There, since the N\'eel (or zigzag) fluctuations are strong, the mixing of different $S^z$-sectors
is not sufficient to mask discrete effects.

\begin{figure}[!t]
\includegraphics[width=0.7\linewidth]{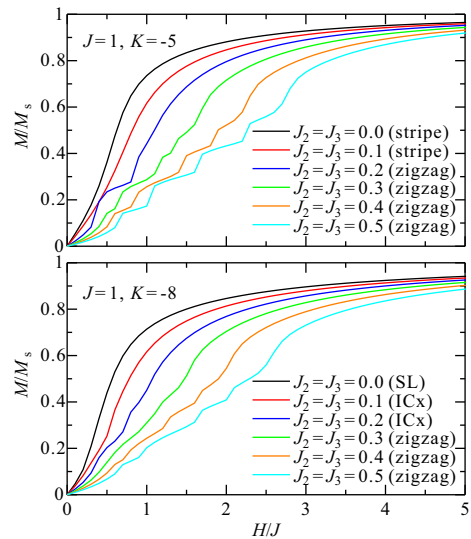}
\caption{
Magnetization curves of the extended Kitaev-Heisenberg model at $J=1$, $K=-5$ (top) and $J=1$, $K=-8$
(bottom), for several values $J_2=J_3$.
For each set $J_2=J_3$, the dominant state at $H=0$ is indicated within parentheses.
The magnetic field is applied along the $z$ direction.
}
\label{fig:S2}
\end{figure}

{\bf Magnetization curves with longer-range interactions}.\,  
We find that $J>0$ and $K<0$ values are required to reproduce the experimental magnetization curves.
Looking in more detail to the dependence on longer-range interactions $J_2$ and $J_3$ is also
instructive.
Magnetization curves at $J=1$, $K=-5$ and $J=1$, $K=-8$ are shown in Fig.~\ref{fig:S2} for several
$J_2\!=\!J_3$ values.
The effect of longer-range interactions seems to be even quantitatively similar for the two
different $K$ values.
As long as $J_2$ and $J_3$ are much smaller than $|J|$ ($J_2, J_3<0.2|J|$), the saturation is
simply shifted to higher field but the overall shape of the magnetization curve is conserved.
On the other hand, for $J_2, J_3>0.3|J|$, the overall shape changes somewhat, approaching that for
the case of $J>0$ and $K>0$.
We thus infer that $J_2$ and $J_3$ must be smaller than $0.3|J|$ to reproduce the experimental
magnetization curves.
Only results for the case of $J_2=J_3$ are shown here for simplicity, since we find that $J_2$ and
$J_3$ have similar effect on the magnetization curves and affect those almost independently.

\begin{figure}[!tb]
\includegraphics[width=0.95\linewidth]{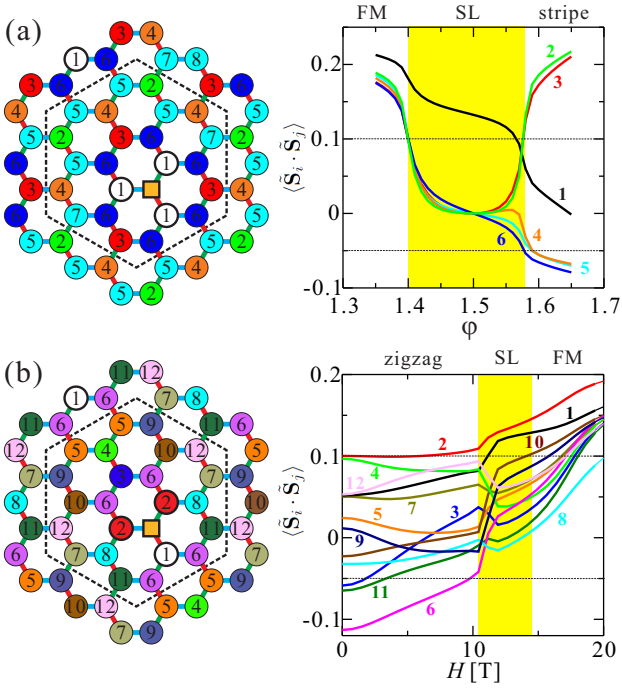}
\caption{
Spin-spin correlation functions $\langle \tilde{\bf{S}}_i\!\cdot\!\tilde{\bf{S}}_j \rangle$ and 
sketch of the periodic clusters used in the ED calculations for (a) a 2D Kitaev-Heisenberg model and
(b) our extended spin model.
The reference site is indicated by a square and the numbers labeling various other sites are in
direct correspondence with the numbered lines in the plots of
$\langle \tilde{\bf{S}}_i\!\cdot\!\tilde{\bf{S}}_j \rangle$.
Yellow windows indicate the Kitaev SL region. 
}
\label{fig:S3}
\end{figure}

{\bf Spin-spin correlations in the spin-liquid phase}.\,  
To describe in more detail the Kitaev SL phase in the intermediate-field region, we calculated the
field-dependent spin-spin correlation functions $\langle \tilde{\bf{S}}_i\!\cdot\!\tilde{\bf{S}}_j \rangle$
and compared them to those of the zero-field Kitaev SL phase of the 2D Kitaev-Heisenberg model on a
honeycomb lattice~\cite{chaloupka10}.
The NN interactions of the Kitaev-Heisenberg model are written as
\[
{\cal H}_{i,j}^{(\gamma)} = 2\bar{K}\, \tilde{S}_i^\gamma \tilde{S}_j^\gamma+\bar{J}\, \tilde{\bf{S}}_i\!\cdot\!\tilde{\bf{S}}_j \,,
\]
where $\gamma(=x,y,z)$ labels the three distinct types of NN bonds in the ``regular'' honeycomb plane.
Following the notation of Ref.~\onlinecite{chaloupka10}, we define the effective parameter
$A=\sqrt{\bar{K}^2+\bar{J}^2}$ and an angle $\varphi$ via $\bar{K}=A\sin \varphi$ and
$\bar{J}=A\cos \varphi$.
In Fig.\,\ref{fig:S3}(a), the spin-spin correlations near the FM Kitaev limit ($\varphi=1.5$) of
the Kitaev-Heisenberg model are plotted, for a 24-site cluster with PBC's.
The Kitaev SL state is characterized by a rapid decay of the spin-spin correlations: in the Kitaev
limit, only the NN correlations are finite and longer-range ones are zero; that is faithfully
reproduced by the 24-site calculations.
Even away from the Kitaev limit, the longer-range (not NN) spin-spin correlations fall within a
narrow range $-0.03 \lesssim \langle \tilde{\bf{S}}_i\!\cdot\!\tilde{\bf{S}}_j \rangle \lesssim 0.1$
in the Kitaev SL phase ($1.40 < \varphi < 1.58$). As seen in Fig.\,\ref{fig:S3}(b), our field-induced SL state exhibits similar features;  the values of longer-range correlations are distributed within a narrow range
$-0.02 \lesssim \langle \tilde{\bf{S}}_i\!\cdot\!\tilde{\bf{S}}_j \rangle \lesssim 0.1$ in the SL phase ($10.8 {\rm T} < H < 14.2 {\rm T}$). In other words, a rapid decay of the spin-spin correlations is seen in our field-induced SL state, at the same level as in the FM Kitaev SL phase of the 2D Kitaev-Heisenberg model.

\bibliography{RuCl3_apr15_RY}

\end{document}